\documentclass[conference]{IEEEtran}
\IEEEoverridecommandlockouts
\usepackage{cite}
\usepackage{amsmath,amssymb,amsfonts}
\usepackage{algorithmic}
\usepackage{graphicx}
\usepackage{textcomp}
\usepackage{xcolor}

\def\BibTeX{{\rm B\kern-.05em{\sc i\kern-.025em b}\kern-.08em
    T\kern-.1667em\lower.7ex\hbox{E}\kern-.125emX}}
\begin{document}

\title{{\color{red} Let's think about the title} \\Scaling the Democratization of Data Intensive Science:\\ Analyzing Petabytes of Climate Data
}
\title{Enabling Scalable and Efficient Analysis of Petascale Climate Data }
\title{Democratizing Scalable Analysis of Petascale Climate Data }
\title{Scalable Ecosystem for Efficient Climate Data Management: Balancing Fidelity and Computational Cost}
\title{Scalable Petascale Climate Data Analysis: Balancing Fidelity and Computational Cost}
\title{Scalable Climate Data Analysis: Balancing Petascale Fidelity and Computational Cost}

\author{\IEEEauthorblockN{Aashish Panta}
\IEEEauthorblockA{\textit{University of Utah} \\
}
\and
\IEEEauthorblockN{Amy Gooch}
\IEEEauthorblockA{\textit{University of Utah}} \\

\and
\IEEEauthorblockN{Giorgio Scorzelli}
\IEEEauthorblockA{\textit{University of Utah}} \\

\and
\IEEEauthorblockN{Michela Taufer}
\IEEEauthorblockA{\textit{U. of Tennessee at Knoxville}} \\
\and
\IEEEauthorblockN{Valerio Pascucci}
\IEEEauthorblockA{\textit{University of Utah} \\
}
}
%AAG: Added this to fix major gap in between title and abstract.. we could do more.
\IEEEaftertitletext{\vspace{-2.5\baselineskip}}

\maketitle

\begin{abstract}

The growing resolution and volume of climate data from remote sensing and simulations pose significant storage, processing, and computational challenges. Traditional compression or subsampling methods often compromise data fidelity, limiting scientific insights. We introduce a scalable ecosystem that integrates hierarchical multiresolution data management, intelligent transmission, and ML-assisted reconstruction to balance accuracy and efficiency. Our approach reduces storage and computational costs by 99\%, lowering expenses from \$100,000 to \$24 while maintaining a Root Mean Square (RMS) error of 1.46 degrees Celsius. Our experimental results confirm that even with significant data reduction, essential features required for accurate climate analysis are preserved. Validated on petascale NASA climate datasets, this solution enables cost-effective, high-fidelity climate analysis for research and decision-making.

\end{abstract}

\section{Problem Overview}

High-resolution climate data from remote sensing and simulations pose significant storage, processing, and computational efficiency challenges. The increasing volume and detail of data incur substantial costs for cloud-based resources and bandwidth. Traditional methods like compression or subsampling reduce costs but often compromise data fidelity, potentially undermining scientific analyses. As datasets grow and analyses become more complex, scalable data management approaches are needed to balance fidelity with computational cost.

% (i)hierarchical, multi-resolution data management and processing techniques
% (ii) Intelligent data transmission techniques and ML-assisted reconstruction methods
% (iii) adaptive fidelity mechanisms
We present a scalable ecosystem that democratizes access to petabytes of scientific data while enabling decision-making. It integrates (i) hierarchical multiresolution data management, (ii) intelligent transmission with ML-assisted reconstruction, and (iii) adaptive fidelity mechanisms. Our approach reduces storage footprint and network transfer times, enhancing scalability for high-performance data analytics. By allowing controlled approximations, our solution maintains analytical fidelity while significantly reducing resource consumption and costs for the scientific community.

We demonstrate the scalability of our ecosystem on the 2.8 PB open-access LLC4320 Ocean climate dataset (ECCO)~\cite{https://doi.org/10.5067/ecl5a-mix44} and 37 TB NEX-GDDP-CMIP6 dataset\cite{nex-gddp}. 
We provide two measures to validate our approach: 

\noindent {\it 1. Accuracy scale} quantifies the trade-off between resolution and data fidelity in climate modeling. While high-resolution data improves accuracy, it also increases computational demands. Our solutions demonstrate effective resolution reduction that preserves meaningful insights, balancing accuracy with resource utilization.

\noindent {\it 2. Efficiency scale} evaluates the balance between data management costs and scientific reliability. As datasets expand, computational expenses become a significant constraint, particularly in cloud environments. We analyze trade-offs between data volume, processing cost, and accuracy to demonstrate how our solution's progressive data reduction substantially decreases data management expenses while preserving scientific integrity and accuracy.

Using our ecosystem, we achieved a 99\% reduction in data usage for 2.8PB of LLC4320 Ocean data, reducing storage, transfer, and computational costs from more \$100,000 to approximately \$24, while maintaining a Root Mean Square (RMS) error of only 1.46°C.

\section{Scalable Data  Ecosystem}
% \subsection{Underlying Concepts and Technologies}
% (i)hierarchical, multi-resolution data management and processing techniques
% (ii) ML-assisted reconstruction methods
% (iii) adaptive fidelity mechanisms

Our scalable solution uses an integrated data ecosystem essential for efficient large-scale data processing. It supports extensive climate datasets through a progressive resolution strategy, machine learning-assisted reconstruction, and adaptive fidelity, which, by acting in concert, enhances storage, transfer, and analytical performance.

% hierarchical, multiresolution data management and processing techniques
At its core, the ecosystem uses {\it hierarchical multiresolution data structures}~\cite{Kumar2014}\cite{10767643}, enabling efficient storage, retrieval, and processing of climate data at various levels of detail. 
Instead of handling a single high-resolution dataset, the ecosystem organizes data into progressively downsampled layers, allowing computations at different resolutions. 
The ecosystem's data management component dynamically revises the progressive resolution by adjusting data fidelity based on task-specific requirements.
For routine analyses, lower-resolution data are processed first, minimizing computational load and I/O overhead. 
When higher accuracy is needed, the management component selectively refines resolution using precomputed multi-resolution representations and ML models to reconstruct fine details from downsampled inputs. Streaming multi-resolution data significantly reduces unnecessary data movement and ensures that high-resolution data is accessed only when required.

%  Intelligent data transmission techniques and ML-assisted reconstruction methods
The ecosystem integrates {\it intelligent data transmission techniques} that prioritize bandwidth-aware data movement to enhance efficiency further. In cloud-based environments, users typically deal with substantial data egress costs. 
Instead of transferring complete datasets, our ecosystem selectively transmits only the required levels of detail, reconstructing full-resolution outputs when necessary using machine learning models. Our strategy significantly reduces network congestion, accelerates computations, and minimizes overall cloud storage and transfer costs.
Another key innovation is the integration of {\it ML-assisted reconstruction} into climate workflows. 
Instead of requiring high-resolution data transfers, the ecosystem employs ML models trained to reconstruct fine details from lower-resolution inputs. 
By learning patterns in climate data, these models enable high-accuracy reconstructions with minimal data movement, achieving a {it Structural Similarity Index (SSIM)} of 0.95 when using half-resolution inputs and maintaining acceptable quality even at extreme downsampling levels such as $\frac{1}{128}$ or $\frac{1}{256}$ of the original data. 

% adaptive fidelity mechanisms

Our ecosystem employs a modular, scalable design {\it adaptable to diverse cloud and HPC environments}. Our ecosystem features a data ingestion and preprocessing pipeline that organizes datasets into a multi-resolution hierarchy, allowing users to access only the required detail level and reducing the I/O burden on large-scale storage systems. Unlike traditional approaches, our framework enables climate scientists to process data at variable resolutions, accessing high-fidelity information only when necessary. The ecosystem's elastic scaling ensures efficient operation across various computational environments, dynamically adapting to available resources to optimize performance and cost. Our practical, deployable solution enables researchers to analyze massive datasets without the prohibitive costs of traditional high-resolution data processing workflows, addressing real-world scientific climate challenges.

\section{Scalability Studies}

% We demonstrate the scalability of our solution in terms of accuracy and computational efficiency within a climate science workflow. Researchers use advanced ocean circulation models combined with historical data from the open-access ECCO consortium dataset to study ocean circulation's impact on climate. However, the dataset's 2.8 PB size imposes significant usage costs, which our solution effectively mitigates. 

We demonstrate the scalability of our solution in terms of accuracy and computational efficiency for the ECCO consortium dataset, which collects simulated data from advanced ocean circulation models combined with recorded data from sensors. The 2.8 PB dataset enables scientists to study the impact of ocean circulation on climate. However, the dataset imposes significant usage costs, which our solution mitigates.

\subsection{Accuracy Scale}

\begin{figure}[!bhtp]
  \centering
    \includegraphics[width=0.9\columnwidth]{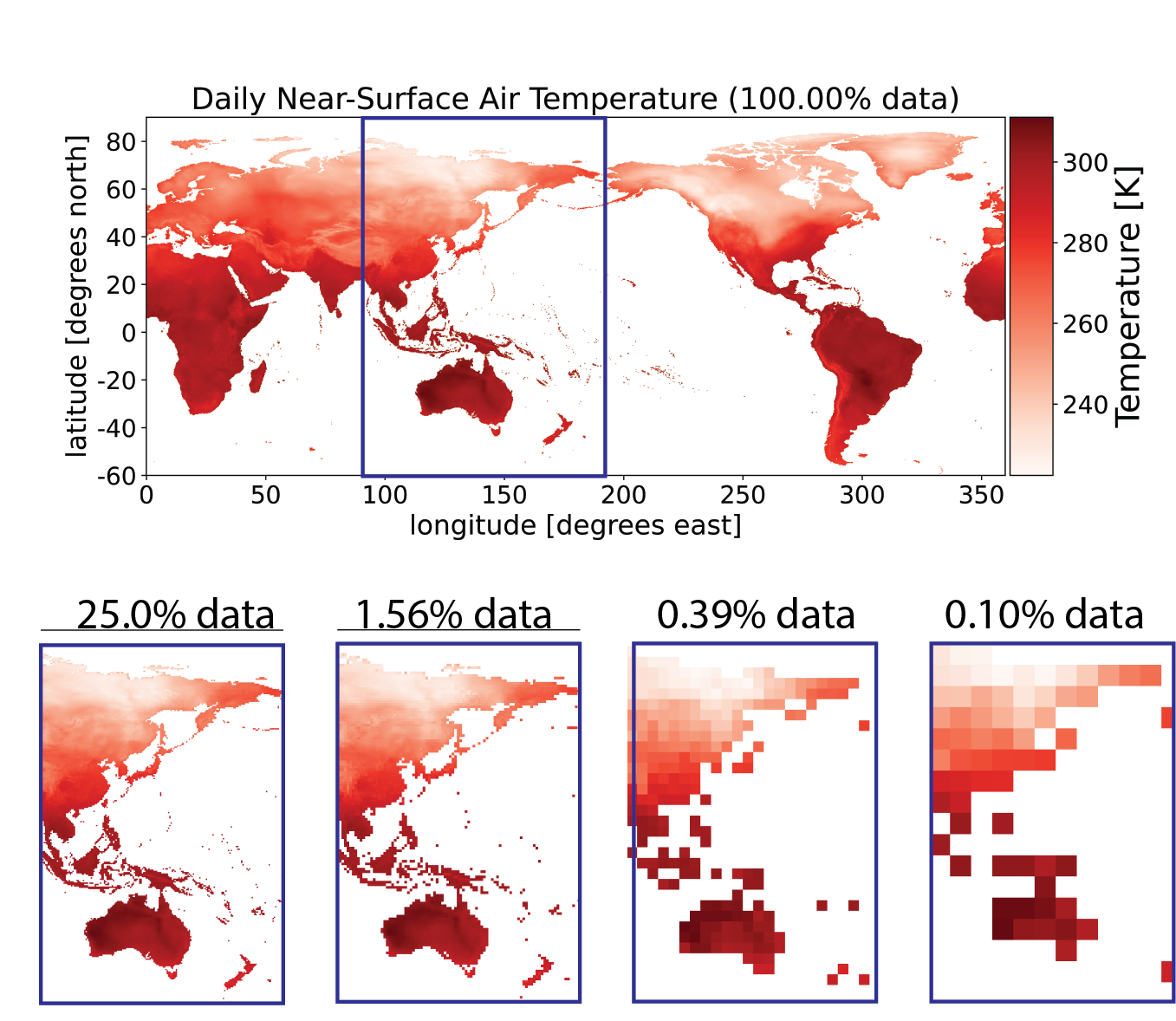} \vspace{-0.5em}
    
  \caption{Progressive downsampling of daily near‐surface air temperature from a fully resolved dataset (top, $100\%$ of the original resolution, with selected region box) with increasingly coarser representations (bottom, from $25.0\%$ to  $0.01\%$ of the data). Each map is colored on the same temperature scale (between 230K and 310K), highlighting how reducing the sampling density preserves broad spatial features but gradually removes finer details, especially in coastal and high‐contrast regions.}
  \label{fig:organized_images}
  \vspace{-1em}
\end{figure}

As the amount of data used in the climate science community grows, efficient data reduction strategies become essential to balance computational cost and data accuracy.
\begin{figure}[!tbp]
    \centering
    \includegraphics[width=0.95\linewidth]{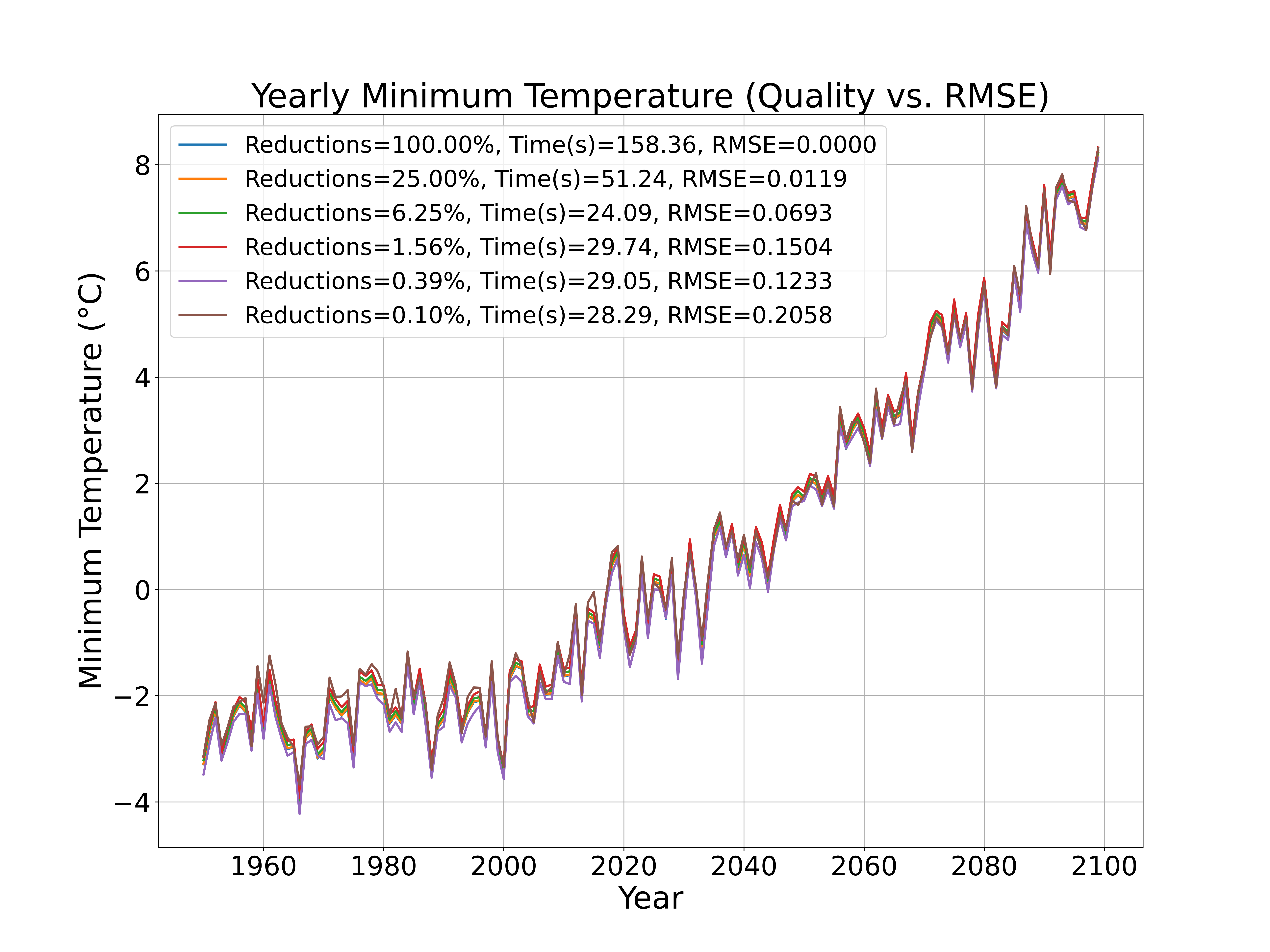}
    \vspace{-1em}
    \caption{Effect of data reduction on climate analysis. Even with significant data downsampling (as low as $0.001\%$ of the original data), the overall trend of the lowest yearly temperature remains well-preserved. The RMSE remains low at moderate downsampling levels, demonstrating that accurate climate insights can be obtained with a fraction of the original data, significantly reducing computational costs and storage requirements.}
    \label{fig:qual_tradeoff}
    \vspace{-1em}
\end{figure}
Figures~\ref{fig:qual_tradeoff} and~\ref{fig:organized_images} demonstrate how our ecosystem enables accurate climate analysis even with significant data reduction, preserving key trends while optimizing computational efficiency.
Specifically, Figure \ref{fig:qual_tradeoff} illustrates the impact of data downsampling on the accuracy of yearly minimum temperature signals from 1950 to 2100. Each colored line corresponds to a different fraction of the original dataset. The legend indicates how much wall‐clock time it took to retrieve the data (in seconds) and the resulting root mean square error (RMSE). 
The full resolution (100\%) data have zero RMSE but takes more than 150 seconds to retrieve. A data set at 25\% reduces the retrieval time to less than 60 seconds with only 0.01°C RMSE. Furthermore, a reduction below 1\% of the data increases the RMSE to 0.15-0.21°C but reduces the retrieval time to around 29 seconds, demonstrating the balance between accuracy and computational efficiency.
Figure~\ref{fig:organized_images} demonstrates the trade-off between data reduction and visual fidelity in large geospatial fields. The figure shows a daily near-surface air temperature dataset progressively downsampled from 100\% (top-left) to 0.098\% (bottom-right) of the original resolution. While coarse resolutions maintain broad temperature gradients, they lose fine spatial details like coastlines and small islands.  
\begin{figure}[!bpt]
    \centering
    \includegraphics[width=0.9\linewidth]{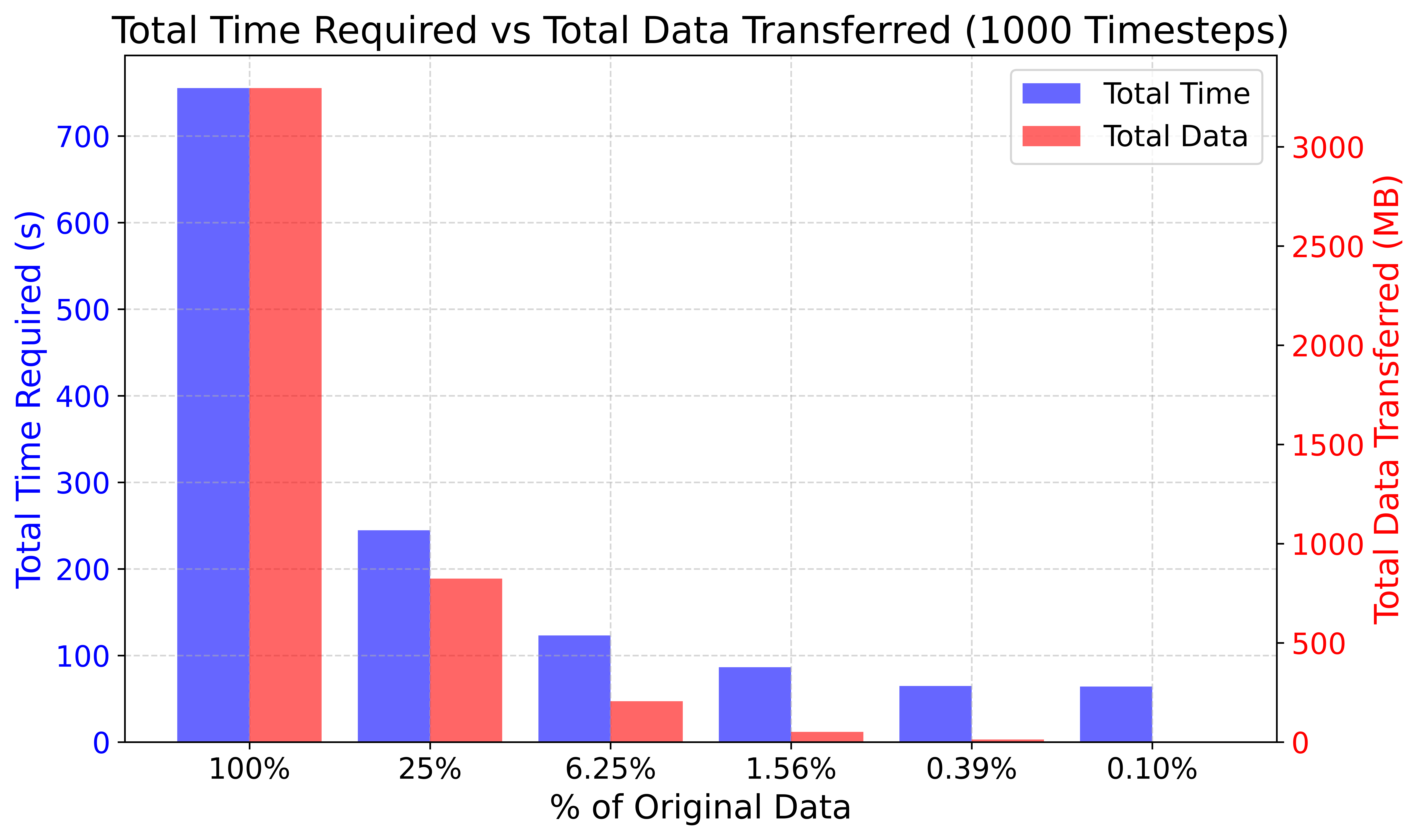}
    \caption{Impact of Data Reduction on Transfer Time and Storage Requirements. The total time required (blue, left axis) and total data transferred (red, right axis) are shown for different quality levels. As the quality level decreases (higher downsampling), both transfer time and data size are significantly reduced, demonstrating the efficiency of progressive data reduction in climate data retrieval.}
    \label{fig:time_v_data}
    \vspace{-2em}
\end{figure}
Our results underscore a clear accuracy-efficiency trade-off in data resolution. Moderate reductions ($25\%$ or $6.25\%$) yield substantial performance gains with minimal accuracy loss, while aggressive downsampling (under 1\%) achieves higher speedups at the cost of higher reconstruction errors. Our approach scales effectively to multi-terabyte or petabyte datasets, allowing proportional management of retrieval times, network bandwidth, and storage overhead through consistent downsampling principles.

Figure~\ref{fig:time_v_data} illustrates the relationship between data retrieval time and total data transferred across different quality levels over 1,000 timesteps. The blue bars represent the total time required for retrieval, while the red bars indicate the total volume of data transferred in megabytes (right y-axis). As expected, reducing the quality level significantly decreases the amount of data that needs to be transferred, leading to faster retrieval times. At full resolution (quality = 0, 100\% of data), the dataset requires the highest storage and retrieval time, with 3.2 GB of data transferred over 755 seconds. As data resolution is reduced, both storage and retrieval time improve, with 1.56\% of the data showing an optimal balance: only 51 MB of data transferred while requiring 70 seconds for retrieval. Beyond this point, additional reductions provide diminishing returns in terms of accuracy.

% \begin{figure*}[!bhtp]
%   \centering
%     \includegraphics[width=0.99\columnwidth]{images/tas_0.png} \vspace{-0.5em}
%     \includegraphics[width=0.99\columnwidth]{images/tas_-2.png}\vspace{-0.5em}\\
%     \includegraphics[width=0.99\columnwidth]{images/tas_-4.png}\vspace{-0.5em}
%     \includegraphics[width=0.99\columnwidth]{images/tas_-6.png}\vspace{-0.5em}\\
%     \includegraphics[width=0.99\columnwidth]{images/tas_-8.png}\vspace{-0.5em}
%     \includegraphics[width=0.99\columnwidth]{images/tas_-10.png}
%   \caption{Progressive downsampling of daily near‐surface air temperature from a fully resolved dataset (top‐left, $100\%$ of the original resolution) to increasingly coarser representations (bottom‐right, just $0.098\%$ of the data). Each map is colored on the same temperature scale (between 230K and 310K), highlighting how reducing the sampling density preserves broad spatial features but gradually removes finer details, especially in coastal and high‐contrast regions.}
%   \label{fig:organized_images}
% \end{figure*}

\subsection{Efficiency Scale}  

As dataset sizes increase, computational costs become a significant constraint, particularly in cloud-based environments. 
\begin{figure}[!pthb]
    \centering
    \includegraphics[width=0.95\linewidth]{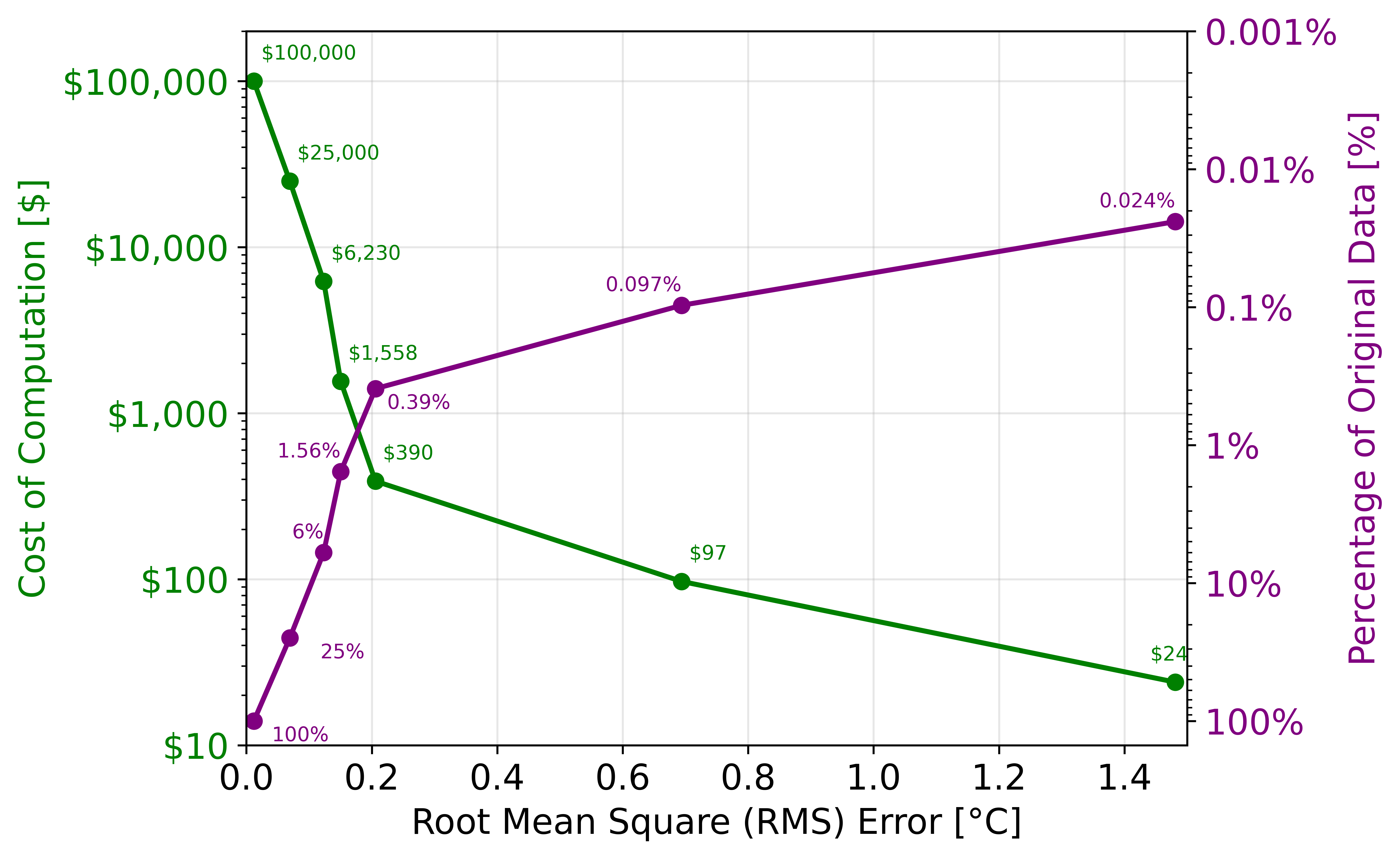}
    \caption{Trade-off between computational cost and data quality. By allowing a controlled RMS error, costs can be reduced dramatically. Our ecosystem enables significant reductions in data utilization (from $100\%$ down to $0.001\%$) while maintaining sufficient accuracy for scientific analysis, demonstrating its scalability and efficiency.}
    \label{fig:cost_vs_qual}
        \vspace{-1em}
\end{figure}

Figure~\ref{fig:cost_vs_qual} demonstrates how our ecosystem enables scalable cost savings and data reduction while preserving sufficient accuracy for scientific analysis, showcasing its scalability and efficiency.
Specifically, Figure~\ref{fig:cost_vs_qual} illustrates how allowing a modest increase in reconstruction error (on the horizontal axis, measured by the RMS error in °C) can drastically reduce both data utilization (right vertical axis) and associated computational costs (left vertical axis, log scale). Near-zero error requires almost full-resolution data, leading to an annual cloud egress cost exceeding $\$100,000$ on data transfer alone, based on AWS egress pricing for US East (Ohio), and outbound data transfer for 105 TB per month\cite{AWS}.  Allowing a slight increase in error drastically reduces costs. For example, tolerating an RMSE of 0.15°C reduces data usage by over 98\%, bringing the cost down to approximately $\$1,500$ per year. At 0.2°C RMSE, data usage drops to just 0.10\% of the original volume, lowering costs to $\$100$ per year. Pushing the reduction further, using just 0.024\% of the original data, makes large-scale climate analysis possible at only $\$24$ per year, but with RMSE of 1.46°C. This tradeoff demonstrates that balancing error tolerance with computational efficiency enables substantial savings while preserving critical climate trends.

Our results highlight efficiency gains in workflows where perfect accuracy is not critical. This cost-benefit analysis reveals that our approach enables cost-effective scalability while maintaining a high degree of accuracy. By adjusting acceptable error thresholds, researchers can scale data handling to the required fidelity, achieving order-of-magnitude reductions in computational effort, network usage, and storage costs.

\section{Impact and Extensibility}

Our approach significantly reduces costs and overhead for massive dataset handling, enabling more frequent high-resolution analyses. The savings can be reallocated to critical needs like computing time or advanced instrumentation. Faster retrieval and processing lead to speedier iteration cycles, allowing researchers to refine models, incorporate new datasets, and rapidly respond to emergent issues. 
This efficiency has policy implications, enabling timely, data-driven insights for decision-makers in climate adaptation, public health, and infrastructure planning. 
The increased scalability democratizes data-intensive research, expanding participation from small laboratories to international collaborations. Ultimately, these advancements bridge the gap between large-scale data generation and practical usage, advancing scientific inquiry, supporting robust policy decisions, and fostering global research collaboration.

\small

\section*{Acknowledgment}
This work was funded in part by NSF OAC award 2138811, NSF
CI CoE Award 2127548, NSF OISE award 2330582, the Advanced Research Projects Agency for Health (ARPA-H) grant no. D24AC00338-00, the Intel oneAPI Centers of Excellence at University of Utah, the NASA AMES cooperative agreements 80NSSC23M0013 and NASA JPL Subcontract No. 1685389. Results presented in this paper were obtained in part using the Chameleon, Cloudlab, CloudBank, Fabric, and ACCESS testbeds supported by the National Science Foundation. This work was performed in part under the auspices of the DoE by LLNL under contract DE-AC52-07NA27344, (LDRD project SI-20-001).
\newpage
% Uncommment these when Valerio accepts the authorship
\bibliographystyle{IEEEtran}
\bibliography{main.bib}

\end{document}